\documentclass[letterpaper, 10 pt, conference]{ieeeconf}  

\IEEEoverridecommandlockouts                              

\usepackage{graphics} 
\usepackage{epsfig} 
\usepackage{mathptmx} 
\usepackage{times} 
\usepackage{amsmath} 
\usepackage{amssymb}  
\usepackage{xcolor}
\usepackage{empheq}
\usepackage{setspace}
\let\labelindent\relax
\usepackage{enumitem}

\usepackage{graphicx} 
\usepackage{subcaption} 

 \renewcommand{\baselinestretch}{0.9}

\title{\LARGE \bf
 SkyGrid: Energy-Flow Optimization at Harmonized Aerial Intersections \thanks{{*This material is based upon work supported by the National Science Foundation under Grant Numbers CNS-2232048, CNS-2204445, and CMMI-2308750.}}}

\author{Sahand Khoshdel$^{1}$, Fatemeh Afghah$^{1}$, Qi Luo$^{2}$ 
\thanks{$^{1}$Sahand Khoshdel and Fatemeh Afghah are with the Holcombe Department of Electrical and Computer Engineering at Clemson University, Clemson, SC, United States
        {\tt\small \{skhoshd,fafghah\}@clemson.edu}.}%
\thanks{$^{2}$Qi Luo is with the Department of Business Analytics at the University of Iowa, Iowa City, IA, United States
        {\tt\small qluo3@uiowa.edu}.}%
}

\begin{document}

\maketitle
\thispagestyle{empty}
\pagestyle{empty}

\begin{abstract}
The rapid evolution of urban air mobility (UAM) is reshaping the future of transportation by integrating aerial vehicles into urban transit systems. 
The design of aerial intersections plays a critical role in the phased development of UAM systems to ensure safe and efficient operations in air corridors. 
This work adapts the concept of rhythmic control of connected and automated vehicles (CAVs) at unsignalized intersections to address complex traffic control problems. 
This control framework assigns UAM vehicles to different movement groups and significantly reduces the computation of routing strategies to avoid conflicts.
In contrast to ground traffic, the objective is to balance three measures: minimizing energy utilization, maximizing intersection flow (throughput), and maintaining safety distances. 
This optimization method dynamically directs traffic with various demands, considering path assignment distributions and segment-level trajectory coefficients for straight and curved paths as control variables. 
To the best of our knowledge, this is the first work to consider a multi-objective optimization approach for unsignalized intersection control in the air and to propose such optimization in a rhythmic control setting with time arrival and UAM operational constraints. A sensitivity analysis with respect to inter-platoon safety and straight/left demand balance demonstrates the effectiveness of our method in handling traffic under various scenarios.
\end{abstract}


\section{Introduction}



UAM has emerged as a promising solution to alleviate traffic congestion and reduce noise and carbon emissions in urban transportation.
As a complementary mode to on-demand ground transportation, UAMs utilize drones and electric vertical take-off and landing (eVTOL) aircraft to transport passengers and goods more rapidly and efficiently.
This emerging mobility service expands the city's transportation options and ultimately integrates into the multi-modal transportation ecosystem to help reduce traffic congestion and improve emergency response times \cite{cohen2021urban}.
However, implementing on-demand UAM services faces challenges such as advanced infrastructure, affordability, and regulatory constraints \cite{pons2022understanding}. 
Designing a reliable end-to-end traffic management system is pivotal for the safe and effective options of UAMs, which
requires considerations across vertiport design, fleet optimization, and demand management \cite{straubinger2020overview}. 


Despite the vastness of the air, UAM can only operate in limited zones due to operational restrictions and noise and privacy concerns. 
As a result, UAM airspace is confined to predefined \textit{air corridors} according to FAA regulations \cite{pons2022understanding}. 
A vast body of literature focusing on geofence repulsion and regulation-compliant routing within these corridors as a separate problem \cite{straubinger2020overview} with the objective of minimizing travel costs between origin-destination (OD) pairs and ensuring safety and geofence compliance.
In contrast, less attention is paid to local intersection controls that 
balances between safety, delay, and energy efficiency.
The required coordination between aerial vehicles in UAM networks is beyond the scale of collision avoidance algorithms because of limitations in real-time communication, data privacy, and urban infrastructure \cite{Xiong2022}.





\begin{figure}[b]
    \vspace{-0.2in}
    \centering
    \includegraphics[width=0.7\linewidth]{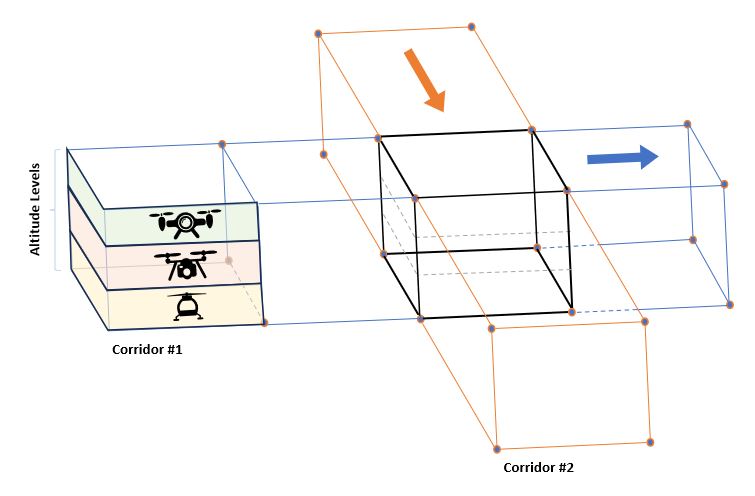}
    \caption{{\small An abstract model for an aerial intersection; the shared airspace of two crossing air corridors.}}
    \label{fig:intersection_corridors}
\vspace{-15pt}
\end{figure}

Aerial intersection are defined as the shared space where two air corridors intersect \cite{nagrare2022multi}. 
Fig. \ref{fig:intersection_corridors} depicts a cube-shaped aerial intersection. 
Routing traffic at such intersections requires reliable and accurate conflict resolution while distributing traffic flow to meet dynamic demand. 
This work is inspired by the concept of unsignalized intersections with CAV applications \cite{talebpour2016, Han2021}.
The standard approach is centralized, real-time control that allocates right-of-way to movement groups dynamically,
requiring a combination of real-time V2V communication and control infrastructure.
These control methods can be challenging to develop and implement. 
More recently, decentralized control with cooperative decisions and communicating with each other directly or through multi-hop vehicular networks has been developed.
Despite this significant improvement without a centralized controller, decentralized control requires ultra-reliable communication between vehicles, real-time cooperation algorithms, and can be challenging to implement in mixed-traffic environments.

Rhythmic control is an alternative to vehicle-based unsignalized control, relying on a communication-efficient, dropout-tolerant approach \cite{chen2021rhythmic}. 
It assigns microsignals to reference nodes within the intersection space and divides the right-of-way for traffic crossing them.
Among works on rhythmic control, strategic holdings allow vehicles to delay their entry into virtual platoons to stabilize traffic flow. 
Several rhythmic control frameworks have been proposed for CAVs, including strategic holdings \cite{lin2021rhythmic} allowing vehicles to delay their entry into virtual platoons to stabilize traffic flow, modular vehicles with rhythmic control \cite{li2023theoretical} enabling vehicles to connect and disconnect to form platoons, and graph coloring control algorithm \cite{abdolmaleki2021unifying} assigning colors to movement groups, allocating right-of-way at nodes.
The advantages of rhythmic control include eliminating waiting time and reducing costs and uncertainties associated with physical signaling, ensuring efficient and predictable flow without heavy real-time communication or sophisticated infrastructure \cite{lin2021rhythmic}. 

Despite many frameworks being proposed to adapt rhythmic control to CAVs, they cannot be directly applied to UAM systems due to distinct operational regulations and settings. 
This is the first work considering the trade-offs between energy consumption, safety requirements, and traffic efficiency in designing UAM intersections.
Our work aims to design energy-aware flow maximization at aerial intersections. First, we build upon rhythmic control to ensure collision-free routing, then maximize intersection flow while minimizing total energy usage through multi-objective optimization on a Pareto frontier. The decision variables are the vehicle-to-path assignment distribution and polynomial coefficients for straight and curved segment trajectories between nodes. We optimize these variables to maximize intersection flow and minimize power consumption.
Our contributions are summarized as follows:

\begin{itemize}[leftmargin=*]
    \item Joint flow-energy optimization of traffic distribution with respect to arrival time constraints of rhythmic control and physical constraints on segment trajectories.
    
    \item Smooth path-level trajectory planning providing flexible intersection routing at harmonized aerial intersections
    
    \item Proposing platoon spacing as a communication-efficient solution to merging harmonized traffic in the intersection.

\end{itemize}


\section{Rhythmic UAM Traffic Control}
\begin{figure}
    \centering
    \includegraphics[width=0.7\linewidth]{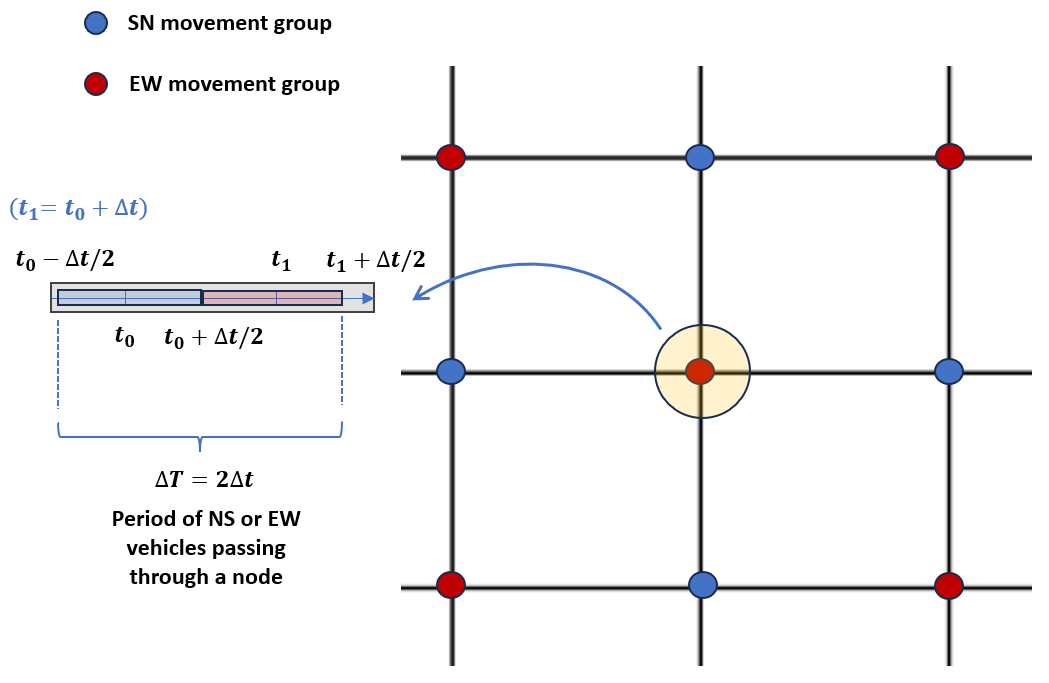}
    \caption{{\small Right-of-way allocation for a sample node. Blue and red colors represent NS and EW movement.}}
    \label{fig:node_allocation}
    \vspace{-15pt}
\end{figure}

\vspace{-0.1cm}
As previously mentioned, rhythmic control is an alternative approach to vehicle-level coordination at unsignalized intersections, which enables no-wait communication-efficient passage \cite{chen2021rhythmic}. Moreover, it was discussed that the intersection of aerial corridors provide the flexibility of using rhythmic control to guide aerial traffic by replacing a physical signal with synchronized arrival times as virtual microsignals. In this method, the intersection flow is divided into two main movement groups; a North-South (NS) group and an East-West (EW) group. For two vehicles travelling towards a node on crossing directions, the arrival time should have the maximum offset possible to minimize the collision risk. Thus, we allocate the right-of-way at any node by dividing the time horizon into two periodic intervals for NS and EW groups, respectively (Fig. \ref{fig:node_allocation}). For example, if it takes $\Delta\,t$ seconds, each, for two vehicles in crossing directions (NS, EW) to move along a straight edge of the intersection graph, the maximum offset of their passing time is also equal to $\Delta\,t$. Thus, every cycle (period) of passing through a node for consecutive vehicles of one movement group will take $\Delta\,T = 2\Delta\,t$, and considering the edge between two nodes in the same direction to be $l_e$ long, there should be a separation of $2\,l_e$ between consecutive vehicles. 
\vspace{0.2cm}

\begin{figure}[h!]
     \centering
     \includegraphics[width=0.7\linewidth]{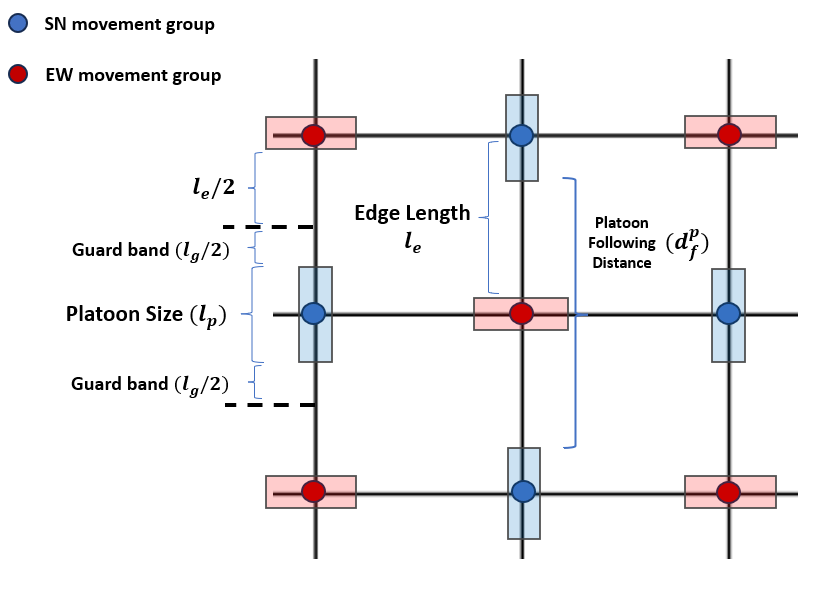}
     \caption{ {\small Right-of-way allocation for NS and EW movement groups considering a guard band of $l_g$ and following distance of $d^{p}_f$. }}
     \label{fig:platoon_allocation}
        \vspace{-15pt}
 \end{figure}

\noindent Considering the passage of only one vehicle in each of the time intervals makes the flow of the intersection be much lower than the capacity and will not support high demand scenarios. To solve this issue, we can extend the passage of one vehicle through a node to a virtual platoon containing multiple vehicles. This way every virtual platoon can cross the intersection at the given interval, although we have to decrease the interval to allow for a guard band ensuring collision free crossing of NS and EW groups, (Fig. \ref{fig:platoon_allocation}). With this design, the large gap between two vehicles moving in the same group ($d_f$) which was previously equal to $2\,l_e$,  is reduced to a smaller gap $ 
 l_e \leq\,d_f \leq 2l_e$. The maximum size of a platoon without guard, for collision-free traffic is equal to $l_p = l_e$, for which the corresponding following distance between consecutive platoons will be minimum. ($\,d_f = l_e$). In this work, we consider a guard band of length $l_g$ equal divided before and after every platoon to account for uncertainties and velocity variations.



\section{System Model}
\vspace{-0.1cm}

\subsection{\textit{\textbf{Intersection Elements}}}
\label{intersection_elements}
 
 
 We model the intersection as a graph $G\,(\mathrm{V},\,\mathrm{E})$, where  $\mathrm{V} = \{v_1, v_2, \cdots, v_N\}$ is the set of $N$ virtual nodes which are equally spaced grid-shaped coordinates that act as routing reference points for the rhythmic control framework, and for 2D routing are all set at a particular altitude level ($v_i = (x_i, y_i, z_i = z_0)$), and $\mathrm{E}$ is the set of edges between these nodes, for which we only consider edges aligned in the NS/EW direction. This results in a  2D planar graph which is a fixed-z layer of the intersection cube (Fig. \ref{fig:intersection_corridors}).
 
\begin{figure} [h]
    \centering
    \includegraphics[width=0.8\linewidth]{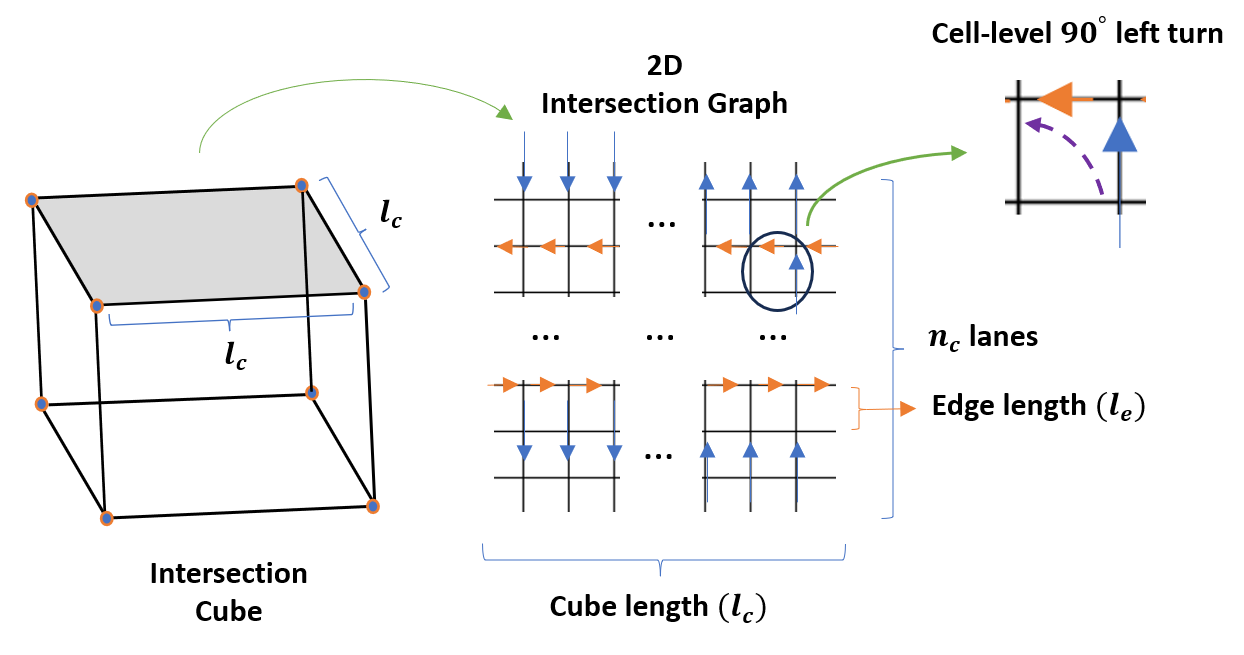}
    \caption{{\small A 4-way Intersection graph, a discretized rhythm reference for rhythmic control on an aerial intersection cube.}}
    \label{fig:intersection_graph}
    \vspace{-5pt}
\end{figure}

 As seen in Fig, \ref{fig:intersection_graph}, for a cube with edge $l_c$, the intersection graph is a $n_c\,\times\,n_c$ grid, for which the edge length can be calculated correspondingly $(l_e = \frac{l_c}{n_c + 1})$. The CAVs can move on each of the $n_c$ lanes of each leg of the intersection. The intersection does allow for no more than one $90^{\circ}$ left turn, and right turns are separately done in the corners of the intersection which we assume are outside the intersection cube that has been modeled. Left turns are completed on quarter circle segments (defined later) between two source and destination nodes of the graph with a radius of $l_e$, meaning they are bounded in a square cell of edge $l_e$ within the grid. $l_e$, which is obviously a divisor of the intersection length $l_c$, is lower-bounded by $l_e^{min}$ that is the minimum curvature radius of the aerial vehicles. The maximum number of lanes in the intersection is determined respectively $(n_c^{max} = \frac{l_c}{l_e^{min}} - 1)$.  Moreover, we assume the number of lanes should always be an even number ($n_c = 2k$) so straight traffic is divided equally between the forward and backward way of each intersection leg.

 We consider two-way traffic on each of the four legs of the intersection, meaning that on each leg, vehicles move from the right-side of the leg, and the intersection is divided to four quadrants with unique combinations of NS/SN and EW/WE traffic. For further reference in the next figures, we only depict the first polar quadrant including SN and EW traffic. We first augment the intersection graph $G\,(\mathrm{V},\,\mathrm{E})$, by adding diagonal edges $\mathrm{E_d}$ to allow for in-cell left turns, resulting in graph $G^{\,'}(\,\mathrm{V}, \mathrm{E}^{\,'});\; \mathrm{E}^{\,'} = \mathrm{E} \cup \mathrm{E_d}$. Edges ($(NW, SW, SE, NE)$) are added in all $1\times1$ cells of polar quadrants $1$-$4$ respectively.

\vspace{2pt}
\noindent \textit{\textbf{Segment}}:\label{def:segments} We define segments $\mathcal{E} = \{e_1, e_2, \cdots, e_{||E^{\,'}||}\}$ as the physical realization of edges $E^{\,'}$ of the augmented graph $G^{\,'}$. A segment uniquely determines the trajectory for vehicles to follow when moving between two consecutive nodes. For future reference straight and curved segments will be referred to as $E_S$, and $E_C$. ($\mathcal{E} = E_S \cup E_C$).

\vspace{2pt}
\noindent \textit{\textbf{Path}}: A Path $p$ is defined as a sequence of $||p||$ segments ($p = \{e_1, e_2, \cdots, e_{||p||}\}$) and connects a source node from one leg of the intersection to a destination node on the opposite leg or the leg on left, satisfying a straight or left-turn demand for a vehicle. For an $n_c\,\times\,n_c$ augmented intersection graph $G'$, there exists $n_c/2$ straight paths and $(n_c/2 - 1)^2$ curved paths (equal to the number of $1\times1$ cells in each quadrant, to ensure safe turns). As previously said, each path can only include zero or one curved segments. 

\vspace{2pt}
\noindent\textit{\textbf{Virtual Platoon}}: A virtual platoon is a hypothetical space bounded around potential vehicles moving on the same segment at a unit time interval (a node beat, $\Delta\,t$). Regardless of the actual number of vehicles $n_v$, the platoon fits the maximum number of vehicles (Fig. \ref{fig:platooning}) and continuously moves in the intersection.



\subsection{\textit{\textbf{Flow Analysis}}}
\label{arrival_analysis}

 Upon arrival at the intersection, and upon arrival at all nodes in the intersection graph, all vehicles must enter at the base speed $V_u$, defined as the velocity for which a vehicle in a platoon traverses a straight edge of the intersection graph, $l_e$, in a unit time interval $\Delta\,t$. ($V_u = \frac{l_e}{\Delta\,t}$) However, vehicles may accelerate/decelerate on segments of the intersection, for which the entry flow can be simply calculated via Eq. \ref{entry_flow_defintition}.  \vspace{-0.3cm}
\begin{equation}
\begin{aligned}
\label{entry_flow_defintition}
f_{ent} &= V_{ent}\,\rho_{ent}\;;\;(\,V_{ent} =V_u = \frac{l_e}{\Delta\;t},\,\rho_{ent} = \frac{n_v}{l_e}\,\frac{1}{K(1+s)}) \\
\Rightarrow &  f_{ent} = \frac{n_v}{\,K\,(1+s)\Delta\,t} = \frac{n_v}{\,4\Delta\,t}
\end{aligned}
\end{equation}
where $n_v$ is the number of the vehicles travelling on an edge/curved segment, $l_e$ the length of straight edge segment, $K$ is the number of harmonized movement groups, which in our case is equal to 2, and s is a binary variable indicating spacing option  for turns (described in subsection \ref{spacing}), which is equal to 1 in our case. The entry flow (Eq. \ref{entry_flow_defintition}) is bounded by the entry capacity (maximum flow) at the intersection, which is controlled by the upper bounds on entry velocity and density. As the entry velocity is set to $V_u = \frac{l_e}{\Delta\,t}$, the capacity (maximum flow) of the intersection entry is controlled by the minimum safe following distance, $d_f^{min}$, the guard band of a platoon, $l_g$, and the length of the vehicles entering the intersection $l_v$, indirectly controlling $n_v^{max}$.

\subsection{\textit{\textbf{Virtual Platoons}}}

We assume that only one platoon (capable of fitting multiple vehicles $n_v$ of a group) passes an intersection node at a unit interval $\Delta\,t$ (time allocated to a NS/EW group at each node). It can potentially contain up to a maximum number of vehicles, $n_v^{max}$ based on its capacity. This upper bound happens when $d_f = d_f^{min}$, and can be found via the following equation:
\begin{equation}
\begin{aligned}
l_p &
= l_e - l_g = n_v\,l_v + n_v\,d_f \Rightarrow \;\mathbf{n_v^{max} = \frac{l_e - l_g}{l_v + d_f^{min}}}
\label{eq: maximum_vehicle_in_platoon}
\end{aligned}
\end{equation}

\begin{figure}
    \centering
    \includegraphics[width=0.7\linewidth]{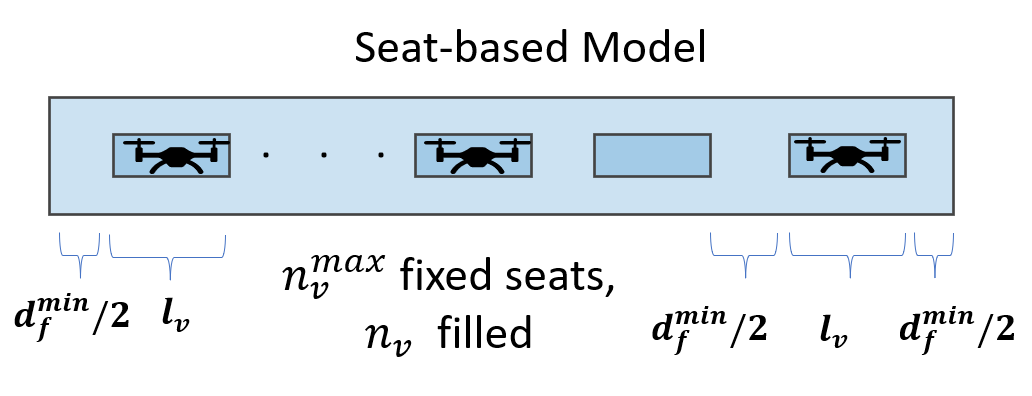}
    \caption{{\small Seat-based platooning model, $l_v$ is the length of the vehicle and $d_f^{min}$ represents the minimum following distance.}}
    \label{fig:platooning}
    \vspace{-15pt}
\end{figure}

These $n_v$ vehicles can be distributed along the platoon in countless ways, but here we consider the most simple design that enables robust routing with respect to V2V communication faults. This model is called a seat-based model (Fig. \ref{fig:platooning}), where the vehicles have corresponding seats which are zones within the platoon, for each vehicle to fit.

\noindent \textbf{\textit{Seating Model}}:
This design assumes the space dedicated to each vehicle on a platoon is fixed and independent of the number of vehicles currently within the platoon (Fig. \ref{fig:platooning}). This distance at all times is equal to $d_f^{min}$ to support the most-congested state where $n_v^{max}$ vehicles are in the platoon. In reality to allow for an uncertainty gap, due to inconsistencies in the CAV controllers or other aerodynamic effects, this $d_f^{min}$ can be set slightly larger than the actual safety distance and act as an intra-platoon headway guard band.  As seen in Fig. \ref{fig:platooning}, as long as vehicles are moving on straight edges with constant velocity $V_u$, their relative speed compared to the following and tailing vehicle will be zero and the distance remains equal to $d_f^{min}$; thus no violation of following distance is made. For when vehicles of varying velocity are moving on straight or curved segments, the velocity profiles, $\dot{x}(t),\dot\theta\,(t)$, are constrained such that all variations still keep $d_f > d_f^{min}$. 


Despite using this seating model strictly limits velocity profiles based on the worst case setting of maximum platoon congestion $n_v = n_v^{max}$, no distance adjustment is needed regardless of which vehicles diverging at which points of the intersection to complete a turn. Thus, communication-tolerant routing can be achieved via such model, as long as vehicles follow the optimized motion profile on curves accurate enough. The coordinates of the $k^{\text{th}}$ vehicle in the platoon (assuming the first vehicle is the one behind all others) can be found via Eq. \ref{eq: vehilce_position_seating}, where $(x_1\,(t), x_0\,(t))$ indicate the first vehicle and platoon start position respectively. 
\begin{equation}
\begin{aligned}
\label{eq: vehilce_position_seating}
x_k\,(t) &= x_1\,(t) + (k-1)(l_v + d_f^{min}) = x_1\,(t) + (k-1)(\frac{l_e - l_g}{n_v}) \\
&= x_0\,(t) + k(l_v + d_f^{min}) = x_0\,(t) + k(\frac{l_e - l_g}{n_v})
\
\end{aligned}
\end{equation}

\begin{figure}
\centering

\begin{subfigure}[t]{\columnwidth} 
    \centering
    \includegraphics[height=1.7in]{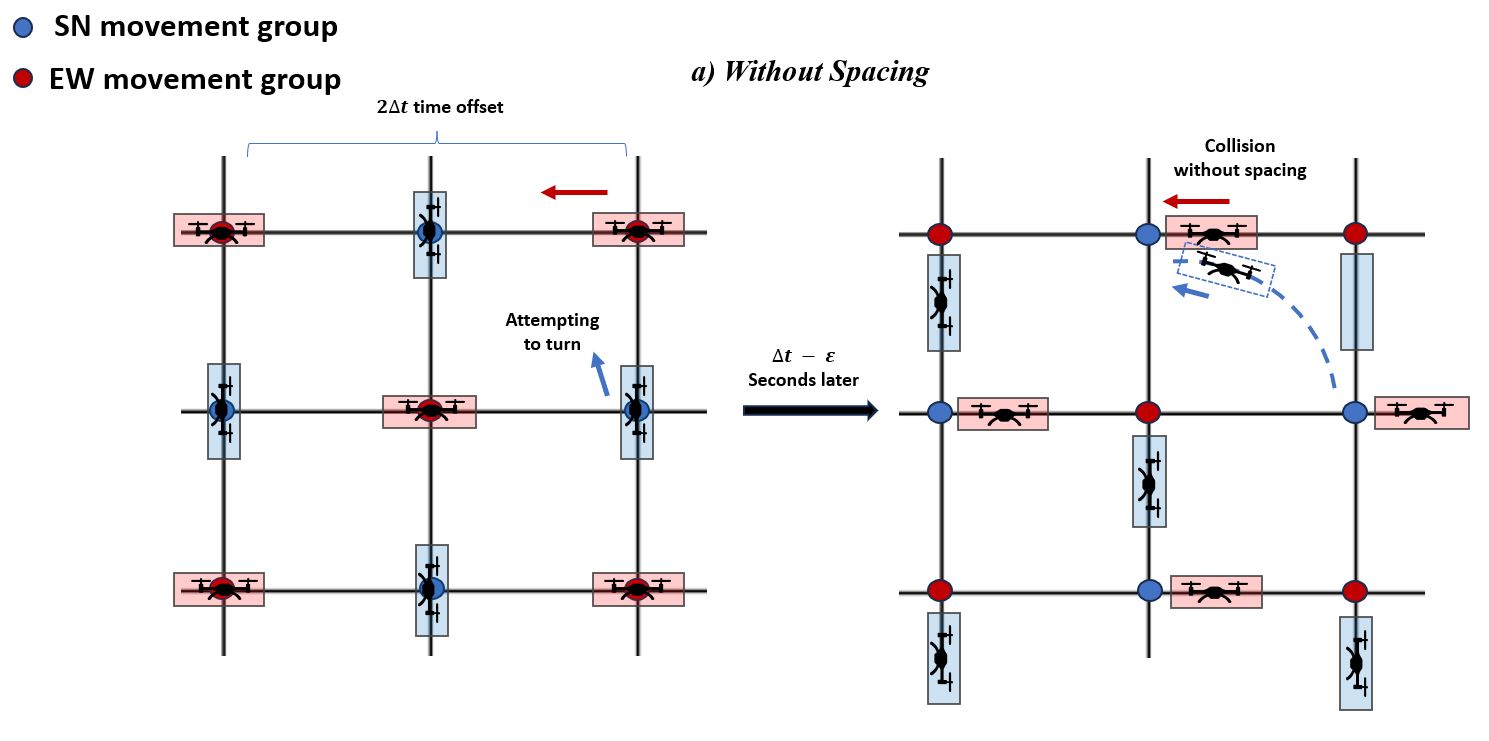} 
    \caption{\small{Collision Happening without spacing}} 
\end{subfigure}%
\vspace{0.2cm}
\begin{subfigure}[b]{\columnwidth} 
    \centering
    \includegraphics[height=1.7in]{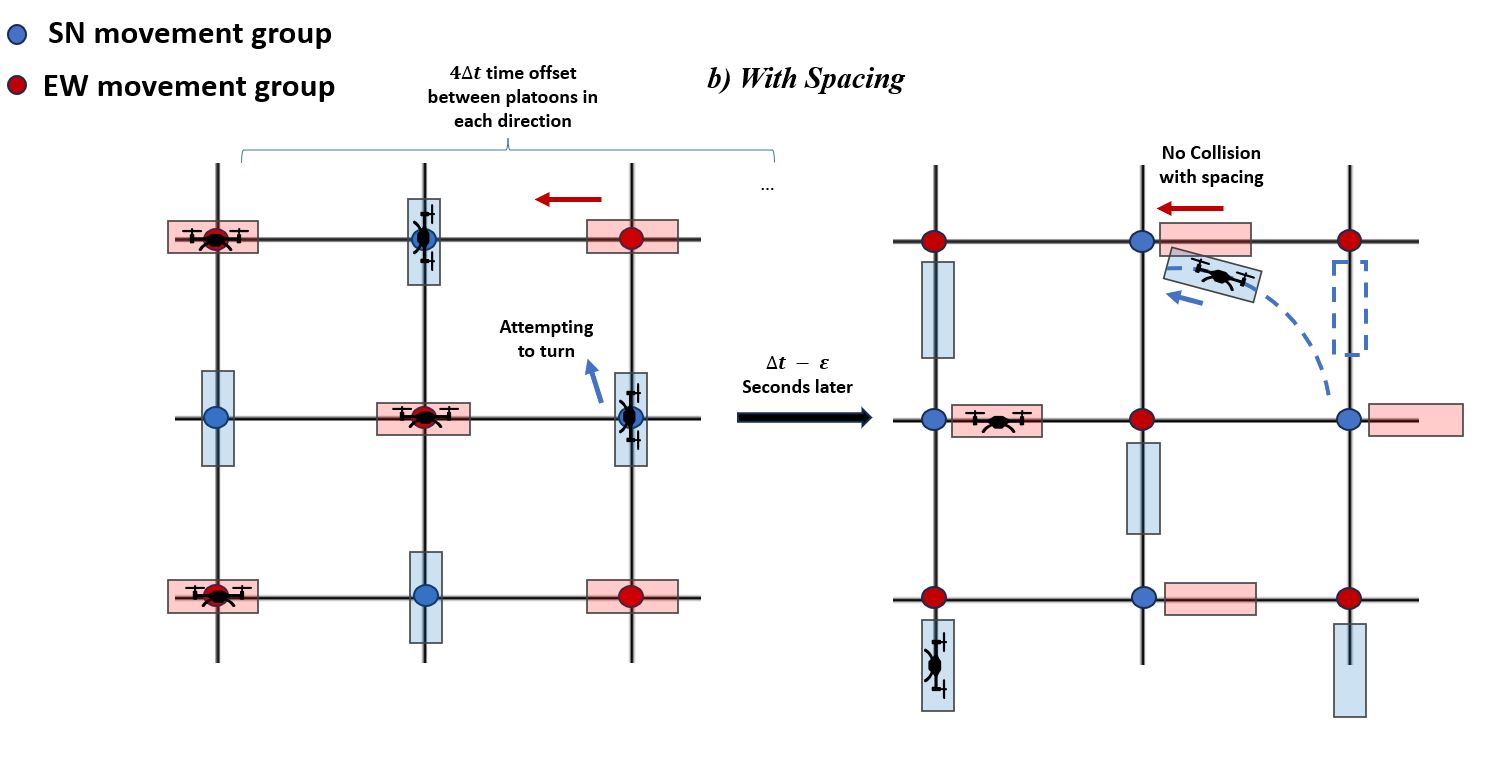} 
    \caption{\small{Collision free merging with spacing}} 
\end{subfigure}

\caption{\small{Comparing two scenarios for merging vehicles at left-turns, with and without spacing. }} 
\label{fig:mering_platoon} 
\vspace{-20pt}
\end{figure}

\vspace{-0.3cm}
\subsection{\textbf{\textit{Platoon Spacing}}}
\label{spacing}
Platoon spacing refers to letting one of two platoons in each direction be empty. The reason spacing is needed for turning within the intersection is to allow for a vehicle (platoon) attempting to turn, to merge into the platoon approaching the destination node in the straight direction without conflict. Without spacing, vehicles of the straight direction (one of the NS or EW platoons) and vehicles attempting to merge, will collide as seen in Fig. \ref{fig:mering_platoon}. The only alternative to this predefined spacing is highly reliable V2V communication and coordination between CAVs to grant turn request in a pairwise manner between the main and the merging platoon, which is beyond the scope of this study. Moreover, platoons with empty vehicle seats grant access to vehicles from merging platoons and only allow for the turning vehicles to fill specific seats. This requires individual intra-platoon distance adjustments, affecting both flow and energy which add to the complexity of the optimization problem. In our model, in addition to platoon spacing, empty and non-empty platoons should be positioned at a specific pattern across the intersection. Each EW platoon ahead of an nonempty SN platoon should be empty, and this should be generalized to every two crossing platoons with a positive $90^\circ$ relative offset on the platoon ahead (merging).






\subsection{\textbf{\textit{Segment Trajectories}}}
\label{segment_trajectories}
The segment trajectories are defined as the path a vehicle traverses on straight and curved segments ($S_{S}, S_{C}$). The 2D trajectory $\{x(t), y(t)\}$ (equivalently in polar coordinates, $\{(r(t), \theta(t)\}$) need to satisfy two initial and two terminal conditions. As seen in Fig. \ref{fig:synchronized_segments}, for straight edges one of $x(t), y(t)$ is always constant along the segment, constraining the other variables motion function with the initial and terminal conditions. For curved edges, we assume trajectories as quarter circles (arcs) from the source node to the destination node. Thus, $\theta\,(t)$ needs to satisfy the conditions, while the radius is constant and equal to the edge length ($r\,(t) = l_e$). With the motion only in one dimension, in the next section we first formulate the segment energy consumption $(E_s. E_c)$ and the segment flow $f_s$, and next calculate the average power $P_{\,tot}$(the average intersection energy consumption in a pattern period) and the intersection flow $f^{\,tot}$ as a function of each segment's flow.

\section{Problem Formulation}

\subsection{\textbf{Segment Endpoint Constraints in Rhythmic Control}}
\label{segment_motion_planning}

Per definition, straight segment trajectories need average velocities equal to $V_u$, to reach the terminal points on appropriate merging time ($\Delta\,t$), while curved segments need a higher average velocity to complete this movement in the same time. Although there exists another platoon spacing, that lets curved segments to merge in $3\Delta\,t$ time but can be proved to increase flow/energy cost. As previously mentioned (\ref{segment_trajectories}), curved segments are modeled as quarter circles in our work, which satisfy the boundary conditions on a radius of $r = l_e$ (Fig. \ref{fig:synchronized_segments}). The path function of degree $K_s$ and $K_c$ for straight and curved segments is defined in Eq. \ref{path_function}, where $a_i, b_i$ are segment coefficients for straight and curved paths and $t_s$ is the entry time. 
\begin{equation}    
\begin{aligned}
&x\,(t - t_s) = \sum_{i = 0}^{K_s}\,a_i\,(t - t_s)^{i}\;;\;\,~K_s\geq\,3\\
&\theta\,(t - t_s) = \sum_{i = 0}^{K_c}\,b_i\,(t - t_s)^{i}\;;\;\,~K_c\geq\,3
\label{path_function}
\end{aligned}
\end{equation}

We consider the path function $x(t), \theta\,(t)$ a polynomial with a minimum degree of 3, satisfying 4 (initial and terminal) constraints in a time interval $\Delta\,t$, which are shown in Eq. \ref{path_function_constraints} and rewritten so the first four coefficients are a function of other coefficients. This way, the optimization variables are reduced to coefficients $(a_4, \cdots a_{K_s})$ and $(b_4, \cdots, b_{K_c})$. 
\begin{equation}    
\begin{aligned}
& x\,(t_s) = 0,\;x\,(t_s + \Delta\,t) = l_e,\;\dot{x}\,(t_s) = \dot{x}\,(t_s + \Delta\,t) = \frac{l_e}{\Delta\,t} \Rightarrow \\
& (a_0, a_1, a_2, a_3) = (0,\,\frac{l_e}{\Delta\,t}, \sum_{i=4}^{K_s}(i-3)a_i\Delta\,t^{i-2}, \sum_{i=4}^{K_s}(2-i)a_i\Delta\,t^{i-3})\\[0.2cm]
& \theta\,(t_s) = 0,\;\theta\,(t_s + \Delta\,t) = \frac{\pi}{2},\;\dot{\theta}\,(t_s) = \dot{\theta}\,(t_s + \Delta\,t) = \frac{1}{\Delta\,t} \Rightarrow \\ & (b_0 = 0, b_1 = 1/\Delta\,t,\,b_2 = \frac{\pi - 3}{{\Delta\,t}^2} + \sum_{i=4}^{K_c}(i-3)b_i\Delta\,t^{i-2}, \\& b_3 = \,\frac{2 - \pi}{{\Delta\,t}^3}\ + \sum_{i=4}^{K_c} (2-i)b_i\Delta\,t^{i-3} 
\label{path_function_constraints}
\end{aligned}
\end{equation}

As a result, for polynomials of degree $K_s$ and $K_c$, there are $K_s - 3$ and $K_c - 3$ degrees of freedom (DoF) respectively. Other than these constraints, some technical constraints should be applied as inequalities over coefficients to ensure physical implementation in the real world. (Eq. \ref{eq:technical_constraints})
\begin{equation}    
\begin{aligned}
 0 &\leq\;\dot{x}\,(t)\;\leq V_{max},\; 0\leq\;\dot{\theta}\,(t)\;\leq V_{max}/l_e  \\
 & x^{(j)}\,(t)\;\leq |x^{(j)}_{max}|,\, \theta^{(j)}\,(t)\;\leq |\theta^{(j)}_{max}|,\;;\;j\geq 2
\label{eq:technical_constraints}
\end{aligned}
\end{equation}

In the equation above, j represents the derivative order, and $\dot{x}, \dot{\theta}$ represent velocity (first-order position derivative). As the j-order derivative of a polynomial of degree $K_s$, is a polynomial of degree $K_s - j$, we can expand the $j$'th derivative of the path function and use it to rewrite Eq. \ref{eq:technical_constraints} as a function on coefficients:
\begin{equation}    
\begin{aligned}
& \overset{(\ref{eq:technical_constraints})}{\Rightarrow} 0  \leq \sum_{i = 1}^{K_s}\,i\,a_i\,(t-t_s)^{i-1}\leq V_{max} \\
 & \overset{(\ref{eq:technical_constraints})}{\Rightarrow} -x^{(j)}_{max} \leq\; \sum_{i = j}^{K_s}\,\frac{i\,!}{(i-j)\,!}\,a_i\,(t-t_s)^{i-j}\;\leq x^{(j)}_{max}\;;\;j\geq 2
\label{eq:position_derivative}
\end{aligned}
\end{equation}

Moreover, as described in Eq. \ref{eq: maximum_vehicle_in_platoon}, the intra-platoon following distance is lower bounded by its minimum ($d_f^{min} = \frac{l_e - l_g}{n_v^{max}} - l_v $). Thus it should be ensured that vehicles moving on a segment do not violate this distance. We model this constraint by calculating the relative position of two consecutive vehicles in a segment at an arbitrary time $t_1$ with respect to each other. As the two vehicles are $\delta\,t = \frac{\delta\,x}{V_u}$ apart upon entering the intersection and they follow the same velocity profile on all edges of a given path, they are always $\frac{d_f}{l_e/\Delta\,t}$ apart in time. The inter-platoon following distance constraint can thus be modeled as seen in Eq. \ref{eq:technical_constraints_2} for straight segment and curved segment only differ in a scaling factor of $1/l_e$ multiplied in the right hand side. 
\begin{equation}    
\begin{aligned}
&x\,(t_0 + \frac{d_f}{l_e/\Delta\,t}) - x\,(t_0) \geq d_f^{min}\;;\;(\, t_s \leq t_0 \leq t_s + \Delta\,t -  \frac{d_f}{l_e/\Delta\,t}) \\
&\overset{Bin.}{\Rightarrow}  \sum_{i = 0}^{K_s}a_i\,[\,\sum_{j=1}^{i}\binom{i}{j}\,{\,(\frac{d_f}{l_e/\Delta\,t}\,)^{j}\;{(t_0 - t_s)}^{i-j}\,}\;] \geq d_f^{min}
\label{eq:technical_constraints_2}
\end{aligned}
\end{equation}

\subsection{\textbf{\textit{Segment Energy Formulation}}}

\begin{figure}
    \centering
    \includegraphics[width=\linewidth]{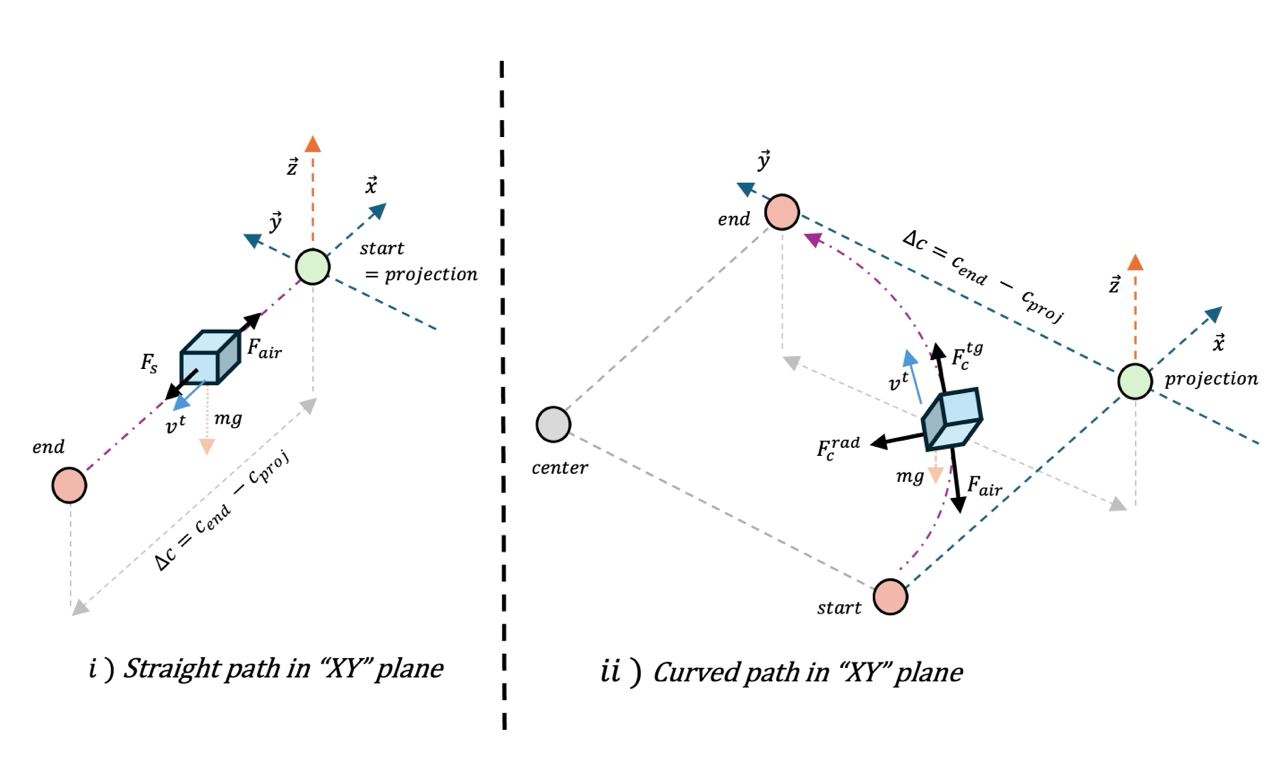}
    \caption{{\small Forces applied to the UAV, while in straight and curved segments of the intersection, in the XY plane.}}
    \label{fig:synchronized_segments}
    \vspace{-10pt}
\end{figure}

The energy required to complete a straight segment, considering a forward acceleration of $a\,(t)$ is as follows (Fig. \ref{fig:synchronized_segments}): \vspace{-0.2cm}
\begin{equation}    
\begin{aligned}
\mathbf{E_s} &= \int_{0}^{\,\Delta\,t}\;P_s\,dt = \int_{0}^{\Delta\,t}\;\vec{F_s}(t)\,\cdot\,\vec{v}(t)\,dt;\; \vec{F_s}(t) = \vec{F}_{air}(t) + m\,a\,(t)\\
&=  (\,\frac{C\,A_{eff}\,\rho_{air}}{2}\,)\,\int_{0}^{ \Delta\,t}|\,\dot{x}^{3}(t)\,|dt + \;m\int_{0}^{\Delta\,t}|\,\dot{x}(t)\,\,\ddot{x}\,(t)|\,\,dt
\label{energy_straight}
\end{aligned}
\end{equation}

For a curved segment, everything is the same as only the tangential component does work, and what makes velocity profiles different are the constraints imposed on them (Eq. \ref{path_function})
\begin{equation}    
\begin{aligned}
\mathbf{E_c} &= \int_{0}^{\,\Delta\,t}\;P_c\,dt = \int_{0}^{\Delta\,t}\;[\,|\,\vec{F_c^{tg}}(t)+\vec{F_c^{rad}}(t)\,|\,]\,\cdot\,\vec{v}(t)\,dt \\
&=  (\,\frac{C\,A_{eff}\,\rho_{air}l_e^{\,3}}{2}\,)\,\int_{0}^{\Delta\,t}|\dot\theta^{3}(t)|\,dt + m\,l_e^2\int_{0}^{\Delta\,t}|\,\dot\theta(t)\ddot{\theta}\,(t)|\,\,dt
\label{energy_curved}
\end{aligned}
\end{equation}

\subsection{\textbf{\textit{Segment Flow Formulation}}}
Flow (throughput) is defined as the number of vehicles passing through a particular point in unit time. To calculate the flow of a segment, we consider the space-average flow along the segment. For straight and curved segments the space-average flow is as follows:
\begin{equation}    
\begin{aligned}
\bar{f_s} &= \int_{0}^{l_e} \rho(x)\,v(x)\,dx =  \frac{n_v}{l_e}\int_{0}^{\Delta\,t} \,v(t)\,\dot{x}\,dt =  \frac{n_v}{l_e}\int_{0}^{\Delta\,t} \,\dot{x}^{\,2}\,dt  \\
\bar{f_c} &= \int_{0}^{\frac{\pi}{2}} \rho(\theta)\,v(\theta)\,d\theta  = \frac{2\,n_v}{\pi\,l_e}\int_{0}^{\Delta\,t} \,v(t)\,\dot{\theta}\,dt = \frac{2\,n_v}{\pi}\int_{0}^{\Delta\,t} \dot{\theta}^2\,dt 
\label{eq:individual_segment_flow}
\end{aligned}
\end{equation}

\subsection{\textbf{\textit{Cost Function Formulation}}}

The objective is to minimize the cost function, which is denoted as $J$, combining the maximization of the total intersection flow and the minimization of the total time-average energy consumption as follows:
\begin{equation}    
\begin{aligned}
& J = \alpha\,{f}_{int} - (1-\alpha){P}_{int};\;{f}_{int} = \phi(\vec{X}_p, \vec{\alpha}, \vec{\beta}),\,{P}_{int} = \psi(\vec{X}_p, \vec{\alpha}, \vec{\beta})\\
& \vec{X}_p = \{x_1, \cdots, x_{||P||}\},\;
\vec{\alpha} = \{a_0, \cdots, a_{K_s}\},\;
\vec{\beta} = \{b_0, \cdots, b_{K_c}\}
\label{eq:cost_function}
\end{aligned}
\end{equation}
The optimization variables are $(\,\vec{X}_p, \vec{\alpha}, \vec{\beta}\,)$, denoting the path assignment distribution, and the straight and curved segment trajectory polynomial coefficients. Formulating the objective $J$ as a function of the control variables can be described in two main steps: 

\noindent1) Rewriting the space-average intersection flow, based on the entry intersection flow, the ratio of CAVs turning left or going straight (known as the demand), and the distribution of assigning CAVs to unique paths satisfying their demands, which can be viewed as finding the function $g$ in $\bar{f}_{int} = \textit{g}\,(\bar{f}_{int}^{ent}, d_s, d_l, x_p)$. \\
2) Calculate the total time-average energy consumption (average power) in the intersection based on the energy consumption of vehicles on each segment and the time-average density of each segment, which can be viewed as finding the function $h$ in $\bar{E}_{int} = \textit{h}\,({E}_{e}, \bar{\rho}_{e})$. \vspace{2pt}

\subsubsection{\textbf{Intersection Flow}}
 We start with assuming incoming traffic towards the intersection with an entry flow $f_{int}^{\,ent}$. We distribute the incoming CAVs between all paths in the intersection that satisfy the CAV demand. $P_s$ and $P_c$ denote the set of straight and curved paths of the intersection, satisfying straight ($d_s$) and left-turn $(d_l)$ demands, respectively. The incoming traffic is divided into straight and left traffic based on their demand and next distributed between paths satisfying that demand as follows:
\begin{align}
f_{int, s}^{\;ent} &= d_s^{\,t}\,f_{int}^{\;ent},\; f_{int, l}^{\;ent} = d_l^{\,t}\,f_{int}^{\;ent}\,;\; (d_s^{\,t} + d_l^{\,t} = 1\,,\,0\leq d_s^{\,t},d_l^{\,t}\leq 1) \nonumber \\
f_{p}^{\;ent} &=  
 x_p\,f_{int}^{\;ent},\sum_{p\in\,P_s}x_p = d_s,\sum_{p\in\,P_l}x_p = d_l\,,\;0 \leq x_p \leq 1;\;\forall p \in P)\label{eq:entry_flow}
\end{align}
\noindent The entry intersection flow and each entry segment flow are limited by the intersection capacity and the straight segment (lane) capacity, respectively, which result in upper bounds on path assignment distribution at the intersection entry for paths that share an entry segment.  
\begin{equation}    
\begin{aligned}
f_{int}^{\;ent} &= f_{int, s}^{\;ent} + f_{int, l}^{\;ent}\\
f_{int}^{\;ent} &= \sum_{e \in \mathcal{E}} f_e \leq n_c\,f_e^{max} \Rightarrow\, \boxed{f_{int}^{\;ent} \leq (\frac{l_c}{l_e} - 1)\,\frac{n_v^{max}}{\Delta\,t}}\\
f_e &= \sum_{p\ni\,e } f_p^{ent}\leq f_e^{max}\,\Rightarrow\, \boxed{\sum_{p\ni\,e}  x_p^{\;ent} \leq \frac{n_v^{max}}{\Delta\,t\,f_{int}};\;\forall e\in \mathcal{E} }
\label{eq:flow_capacity}
\end{aligned}
\end{equation}

After discussing the limits on path distribution and intersection entry flow in Eq. \ref{eq:entry_flow},\ref{eq:flow_capacity}, each segment's flow can be calculated by summing up the flow of all paths that include it. Considering each path with $x_p\,f_p^{ent}$ vehicles flowing through it for a $4\Delta\,t$ period cycle, the average flow of a segment within the path and not common with other paths ($f_e^{p}$) can be found via integrating its flow over a period as seen in Eq. \ref{eq:segment_flow}. Here,  $\bar{\rho}_e\,(t)$, $\bar{V}_e\,(t)$ denote the space-average density and velocity of vehicles over a segment at time $t$. 
\begin{equation}    
\begin{aligned}
f_e^{p} = E_{T}\,(\bar{f}_e) = \frac{\int_{T}\,\bar{f}_e\,(t)}{T} &= \frac{\int_{T}\,\bar{\rho}_e\,(t)\,\bar{V}_e\,(t)}{4\Delta\,t} = \frac{\bar{V}_e\,\int_{4\Delta\,t}\,\bar{\rho}_e\,(t)\,}{4\Delta\,t} \\
\bar{\rho}_e\,(t) = \frac{n_v^{e}(t)}{l_s},\;\bar{V}_e\,(t) &= \frac{\int\,V(t)\frac{ds}{dt}\,dt}{\int\,ds} = \int_{0}^{\Delta\,t}\,V^2(t)\,dt/l_s
\label{eq:segment_flow}
\end{aligned}
\end{equation}

\noindent The space average velocity of a segment is an integration over space, which can be converted to an integration over time using the path function. As a result, $\bar{V}_e\,(t)$ is independent of time and can be factored out of the integral, as the velocity profile of a segment remains the same for any cycle period $4\Delta\,t$. On the other hand, the space-average density is the number of vehicles on the whole segment at time $t$ divided by the segment length. The space-average density varies as vehicles enter and exit the segment. Therefore, we break up the integral into a steady interval ($\Delta\,t_{ss}$) where all vehicles are enclosed to the segment and a transient interval ($\Delta\,t_{tr}$) where vehicles are entering/exiting the segment. The length of the steady interval is independent of the path flow in a seating-based platooning. In contrast, the integration of the transient interval can be proven independent of the vehicle entry/exit order. That being said, the segment flow is obtained as a function of the path entry flow as:
\begin{equation}    
\begin{aligned}
f_e^{p} = \frac{\bar{V}_e\,}{4\Delta\,t\,l_s}\,\int_{4\Delta\,t}\,n_v^{e}(t)\,dt &= \frac{\bar{V}_e\,}{4\Delta\,t\,l_s}\,[\,\int_{\Delta\,t_{tr}}\,n_v^{e}(t)\,dt\,+\,\int_{\Delta\,t_{ss}}\,n_v^{e}(t)\,dt]  \\
 \Rightarrow f_e &= \frac{\bar{V}_e\,}{l_s}\,(\frac{2\,n_v^{max}-1}{n_v^{max}})(\frac{l_e - l_g}{l_e}\,){\Delta\,t}\,f_p^{ent}\, 
\label{eq:segment_flow_2}
\end{aligned}
\end{equation}

\noindent The equation above describes the ratio of a segment flow to an entry path flow. This ratio is different for curved and straight segments and can be further expanded by considering the segment utilization ratio as a constant controlled by the intersection design ($X_e = l_e - l_g$). This results in Eq. \ref{eq:segment_flow_3}, which describes the ratio as a space-average velocity factor ($\bar{V}_e/l_sV_u$), a segment utilization factor ($X_e$) and a segment capacity factor ($2 - 1/n_v^{max}$).
\begin{equation}    
\begin{aligned}
   f_e &= \frac{\bar{V}_e\,X_e}{V_u\,l_s}\, (2 - 1/{n_v^{max}}){f_p^{\,ent}}
    ;\;\bar{V}_e\, = \left\{\begin{matrix}
\int_{0}^{\Delta\,t}\,\dot{x}^2(t)\,dt/l_e^{3};e\in\,E_s\\[0.2cm]\frac{4}{\pi^2}\int_{0}^{\Delta\,t}\,\dot{\theta}^2(t)\,dt/l_e;e\in\,E_c
\end{matrix}\right.
\label{eq:segment_flow_3}
\end{aligned}
\end{equation} 

\noindent As the only terms dependent on the optimization coefficients are the average segment velocity and the path entry flow, we factor out all other terms as a unified design term $X_{int} = (X_e/V_u)\,(2 - 1/n_v^{max})$, and expand the flow equation once more, resulting in Eq. \ref{eq:segment_flow_4}.
\begin{equation}    
\begin{aligned}
    &\;f_e^{p} = X_{int}\frac{\bar{V}_e}{l_s}\,f_p^{\,ent} =  f_{int}^{ent}\,X_{int}\frac{\bar{V}_e}{l_s}\,x_p\,\\[-0.4cm]
\label{eq:segment_flow_4}
\end{aligned}
\end{equation}

\noindent Finally, each path's space-average flow can be calculated using a weighted space-average of all segment flows within the path, where the weights are the ratio of segment to path length. The average intersection flow can finally be obtained by summing all the individual path flows within the intersection. 
\begin{align}
\quad\Rightarrow  f_{int} &= \sum_{p \in P}\,f_{p} =  \sum_{p \in P} (\sum_{e\in\,p}\; \frac{{l_e\,}}{\sum_{e\in\,p}\; {l_e}}f_{e}^{p}) 
\label{eq:total_intersection_flow}
\end{align}

The segment weights can be rewritten as the number of curved and straight segments in a path and their length. Using Eq. \ref{eq:segment_flow_4} to substitute for $f_e^{p}$, the intersection flow in Eq. \ref{eq:total_intersection_flow} can be written as a function of average velocity, path distribution and energy of straight and curved segments (Eq. \ref{eq:total_intersection_flow_2}). $n_s^p$ shows the number of straight segments in a curved path. $\bar{V}_{e_s}, \bar{V}_{e_c}$ show space average segment velocities on straight and curved paths respectively.
\begin{equation}    
\begin{aligned}
f_{int}& = f_{int}^{ent}\,X_{int}\sum_{p \in {P}} \frac{x_{p}}{l_p}\sum_{e \in p}\bar{V}_e  \\ 
& = f_{int}^{ent}\,X_{int}\sum_{p \in {P}} \frac{x_{p}(n^p_s\bar{V}_{e_s}+ n^p_c\bar{V}_{e_c})}{(n^p_s+(\pi/2)\,n^p_c)l_e} \\
& =  f_{int}^{ent}\,\Delta\,t\, (\frac{l_e - l_g}{l_e})\,(2 - \frac{1}{n_v^{max}})\;[\frac{\bar{V}_{e_s}}{l_e}\,d_s    +  \sum_{p \in {P_{c}}}\,\frac{x_{p}(n^p_s\bar{V}_{e_s}+ \bar{V}_{e_c})}{(n^p_s+\pi/2\,)l_e} ] \\ &= \boxed{\rho_{int}^{ent}\, (2 - \frac{2l_g + d_f^{min}}{l_e})\;[\bar{V}_{e_s}\,d_s    +  \sum_{p \in {P_{c}}}x_p\,\frac{n^p_s\bar{V}_{e_s}+ \bar{V}_{e_c}}{n^p_s+\pi/2\,} ]} 
\label{eq:total_intersection_flow_2}
\end{aligned}
\end{equation}

As seen in Eq. \ref{eq:total_intersection_flow_2}, the intersection flow does not depend on the distribution of vehicles among the possible straight paths, as all have only straight segments, which makes their space-average flow equal. Moreover, the entry density upper bound is controlled by the number of entry lanes $n_c$ and the maximum number of vehicles $n_v^{max}$ in each, which itself is controlled by $l_g, d_f^{min}$ safety parameters. ($\rho_{int}^{ent} \leq \frac{n_c\,(l_e - l_g)}{8\,d_f^{min}}$) 

\subsubsection{\textbf{Intersection Energy (Average Power)}}
To calculate the time-average energy (average power), we consider a full pattern cycle of $T = 4\Delta\,t$ and calculate the energy of vehicles in each path based on the energy of a straight and curved segment. The average consumed power of an intersection is shown in Eq. \ref{eq:average_power_2}. $||p||$ denotes the path length. 
\begin{align}
\label{eq:average_power_2}
P_{int} &= \frac{E_{int}^{tot}}{4\Delta\,t} = \frac{\sum_{p \in P}{E_p^{tot}}}{4\Delta\,t} =  \frac{\sum_{p \in P}\sum_{veh. \in p}{E_p^{veh}}}{4\Delta\,t} \\ \nonumber 
\Rightarrow P_{int}&=  \rho_{int}^{ent}\,V_u\,[\,\sum_{p \in P_c} x_p(||p||E_s + E_c - E_s) + \,\sum_{p \in P_s} x_p(||p||E_s)] \\ \nonumber
&= \boxed{\rho_{int}^{ent}\,V_u\,[\,\sum_{p \in P_c} x_p(||p||E_s + E_c - E_s) + \,d_s(n_c+1)\,E_s]}
\end{align}

As seen in this equation, likewise the intersection flow, the average power does not depend on the path assignment distribution in straight paths. Finally the total objective function can be formulated as follows:
\begin{equation}    
\begin{aligned}
J & = \rho_{int}^{ent} [\alpha\, (2 - \frac{2l_g + d_f^{min}}{l_e})\;(\sum_{p \in {P_{c}}}x_p\,\frac{(||p||-1)\bar{V}_{e_s}+ \bar{V}_{e_c}}{(||p||-1)+\pi/2\,} + \\& + \bar{V}_{e_s}\,  \sum_{p \in {P_{s}}}x_p)
+ (1-\alpha)\,V_u\,(\,\sum_{p \in P_c} x_p(||p||E_s + E_c - E_s) \\& + \sum_{p \in P_s}x_p\,(n_c+1)\,E_s)] \\
 \bar{V}_{e_s} & = \int_{0}^ {\Delta\,t} (\sum_{i = 1}^{K_s}\,i\,a_i\,t^{i-1})^2\,dt,\;
  \bar{V}_{e_c}  = l_e\int_{0}^ {\Delta\,t} (\sum_{i = 1}^{K_c}\,i\,b_i\,t^{i-1})^2\,dt  \\ 
E_s &=    (\,\frac{C\,A_{eff}\,\rho_{air}}{2}\,)\,\int_{0}^{ \Delta\,t}\;(\sum_{i = 1}^{K_s}\,i\,a_i\,t^{i-1})^{3}\,dt \\ & +  \;m\int_{0}^{\Delta\,t}\,(\sum_{i = 1}^{K_s}\,i\,a_i\,t^{i-1})\,\,(\sum_{i = 2}^{K_s}\,i(i-1)\,a_i\,t^{i-2})\,\,dt \\
E_c &=    (\,\frac{C\,A_{eff}\,\rho_{air}l_e^3}{2}\,)\,\int_{0}^{ \Delta\,t}\;(\sum_{i = 1}^{K_s}\,i\,b_i\,t^{i-1})^{3}\,dt \\ & +  \;ml_e^2\int_{0}^{\Delta\,t}\,(\sum_{i = 1}^{K_s}\,i\,b_i\,t^{i-1})\,\,(\sum_{i = 2}^{K_s}\,i(i-1)\,b_i\,t^{i-2})\,\,dt 
\label{eq:total_objective}
\end{aligned}
\end{equation}

The optimization problem is as maximizing J, with respect to the path assignment distribution, and path polynomials as seen in Eq. \ref{eq:optimization_problem}, while satisfying the equality constraint of arrival times on path polynomial coefficients (Eq. \ref{path_function_constraints}), the path share probability constraint (Eq. \ref{eq:entry_flow}), and the inequality constrains on path derivatives (Eq.\,\ref{eq:position_derivative}), the minimum intra-platoon following distance (Eq.\,\ref{eq:technical_constraints_2}), and the edge capacity constraint (Eq. \ref{eq:flow_capacity}).
\begin{equation}
\begin{aligned}
&Max \; J\;\text{w.r.t.}(x_{1}, \cdots, x_{||P_c||+||P_s||}), (a_0, \cdots, a_{K_s}), (b_0, \cdots, b_{K_c}) \\
&\text{s.t.} \quad (\text{Eq.}~\ref{path_function_constraints},~\ref{eq:position_derivative},~\ref{eq:technical_constraints_2},~\ref{eq:entry_flow}, ~\text{and}~\ref{eq:flow_capacity})
\end{aligned}
\label{eq:optimization_problem}
\end{equation}

\section{Evaluation}

The multivariate constrained optimization problem was approached using the COBYLA solver in the SciPy python library, with initializing the path assignment distribution to a uniform distribution over all straight/curved paths respecting their demands  ($x_p^{\,0} = 
\frac{d_s}{||P_s||};\;p \in P_s, x_p^{\,0}  = \frac{1-d_s}{||P_c||};\;p \in P_c
$). Intersection settings are initialized to ($n_c = 6, l_e = 10, \Delta\,t = 1, \rho_{int}^{ent} = 3/l_e, l_v = 0.5, d_{f_{min}} = 1.5$), with degree $K_s = K_c = 4$ segment trajectory polynomials, and $a_4 = b_4 = 0$.

\subsection{Optimization}

\begin{figure}
    \centering
    \subfloat[][]{\includegraphics[width=0.33\columnwidth]{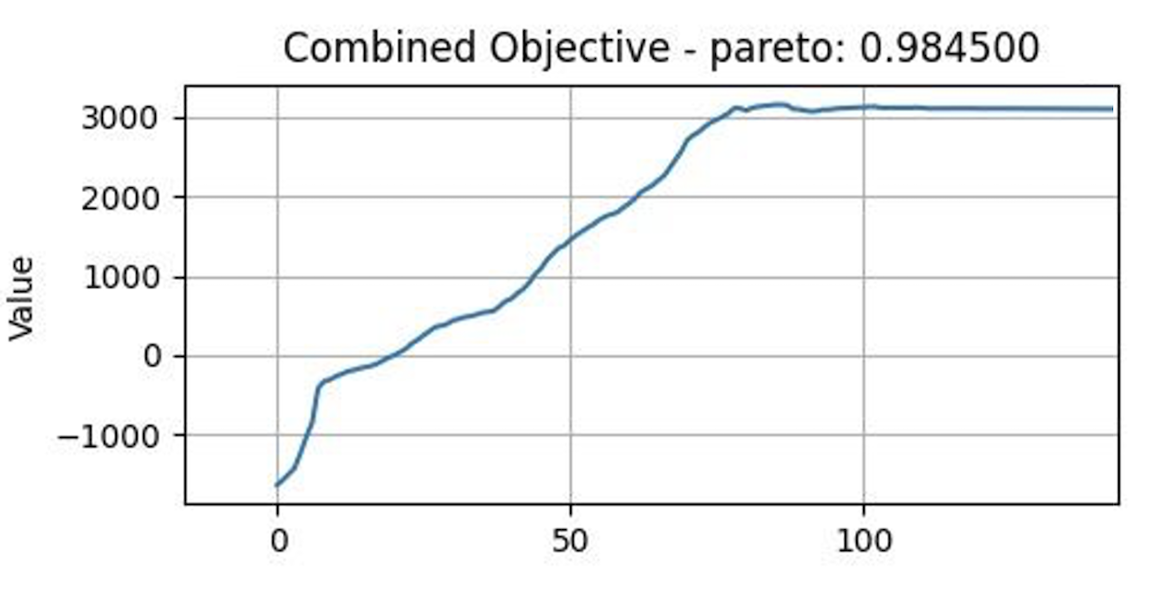}} 
    \subfloat[][]{\includegraphics[width=0.33\columnwidth]{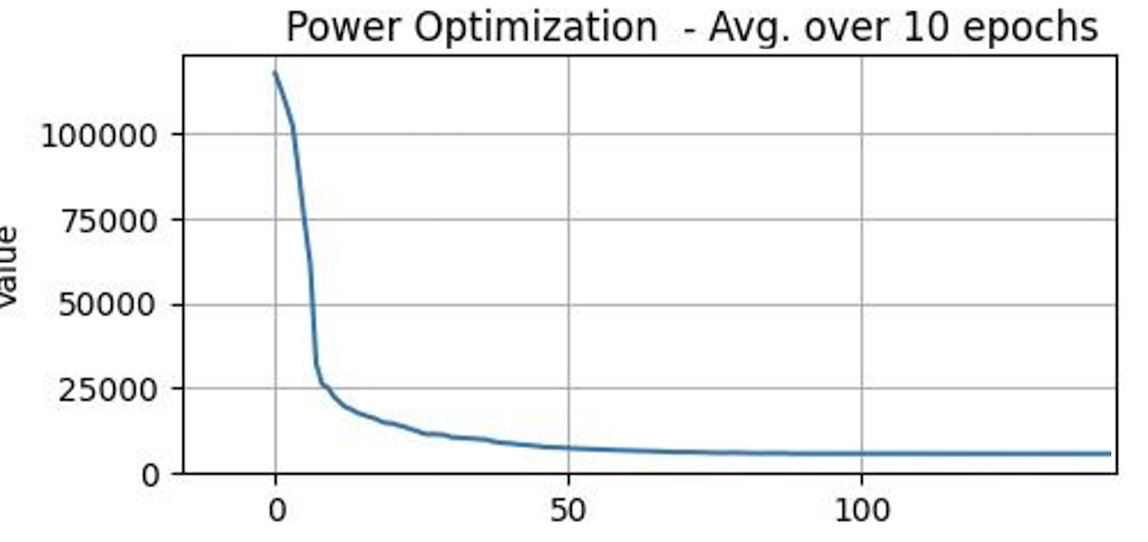}} 
    \subfloat[][]{\includegraphics[width=0.33\columnwidth]{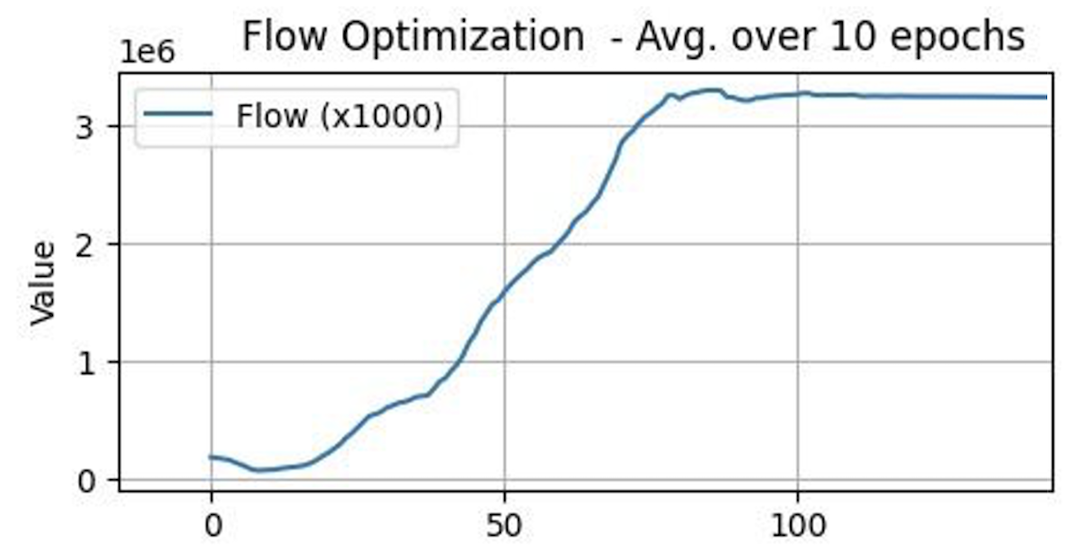}}
    \caption{{\small Optimizing the joint objective - the solver converges after 100 iterations, with increasing flow and decreasing power.}}
    \label{fig:objective_convergence}
\end{figure}

\begin{figure}
    \centering
    \includegraphics[width=0.7\linewidth]{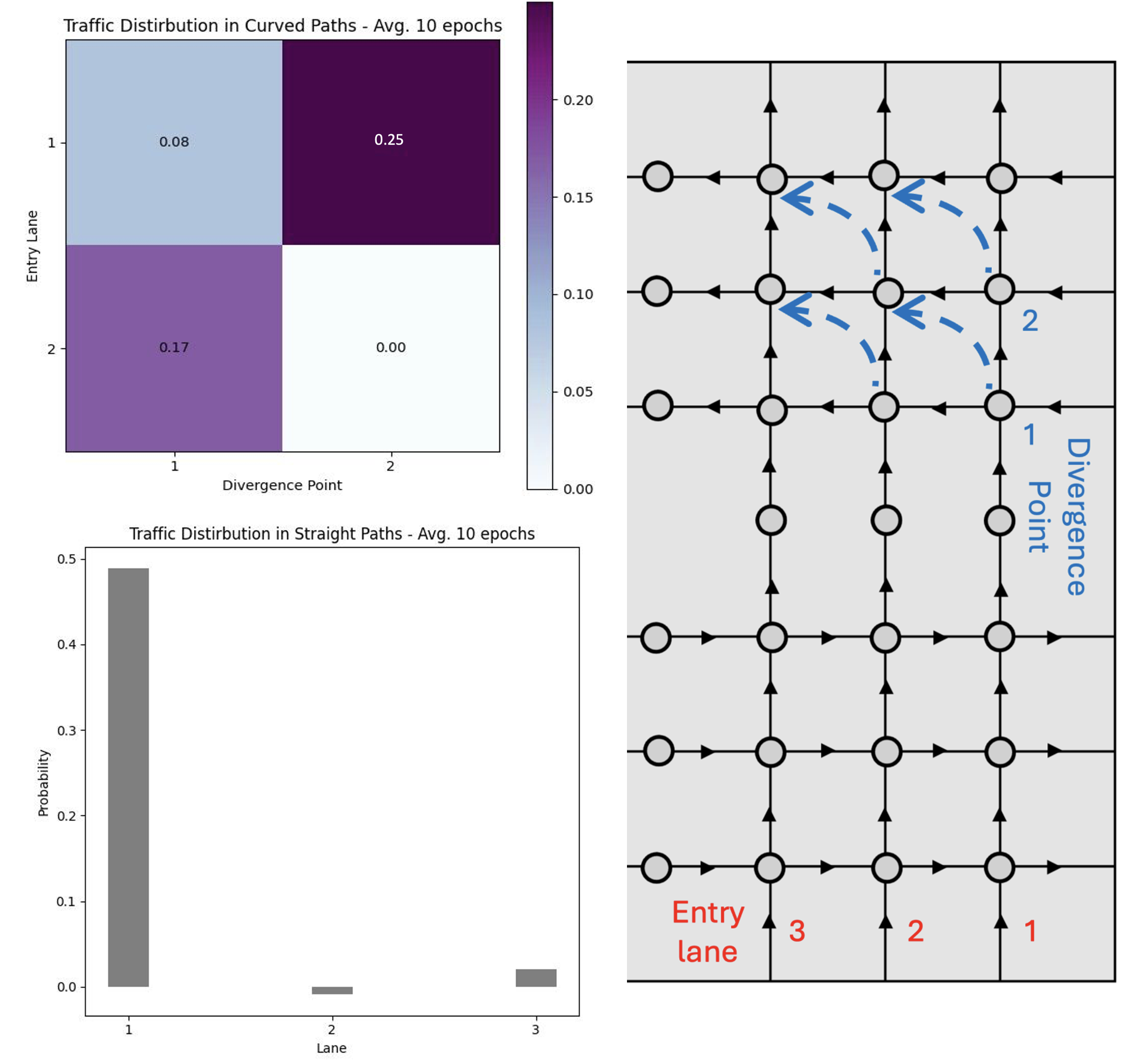}
    \caption{{\small Path assignment distribution on straight and curved paths, shown in a bar chart and heatmap respectively. On the right, a sample 6x6 intersection is shown with entry lane and divergence point concepts depicted. }}
    \label{fig:optimal_variables}
     \vspace{-5pt}
\end{figure}

Under the aforementioned settings, it is observed that intersection flow and power have very different scales, resulting in an optimal combination weight of $\alpha = 0.9845$ as seen in Fig. \ref{fig:objective_convergence}. Convergence is reached within 100 iterations, when averages over 10 epochs, while decreasing power and increasing flow is observed. Fig. \ref{fig:optimal_variables} depicts the path assignment distribution at the optimally point of this sample setting. In a heatmap, the path distribution of curved paths, which are each separated by their entry lane and their turning (divergence) point along the lane, is shown. In the intersections used in our experiments, the leftmost lane (lane 3 here) does not allow for turns and is dedicated to straight traffic, while all other lanes allow for turns and throughs(straight traffic). In each lane the turning points are labeled from 1, starting from the earliest turning point. 

\subsection{Inter-platoon Safety Analysis}
\begin{figure}
    \centering
    \includegraphics[width=1\linewidth]{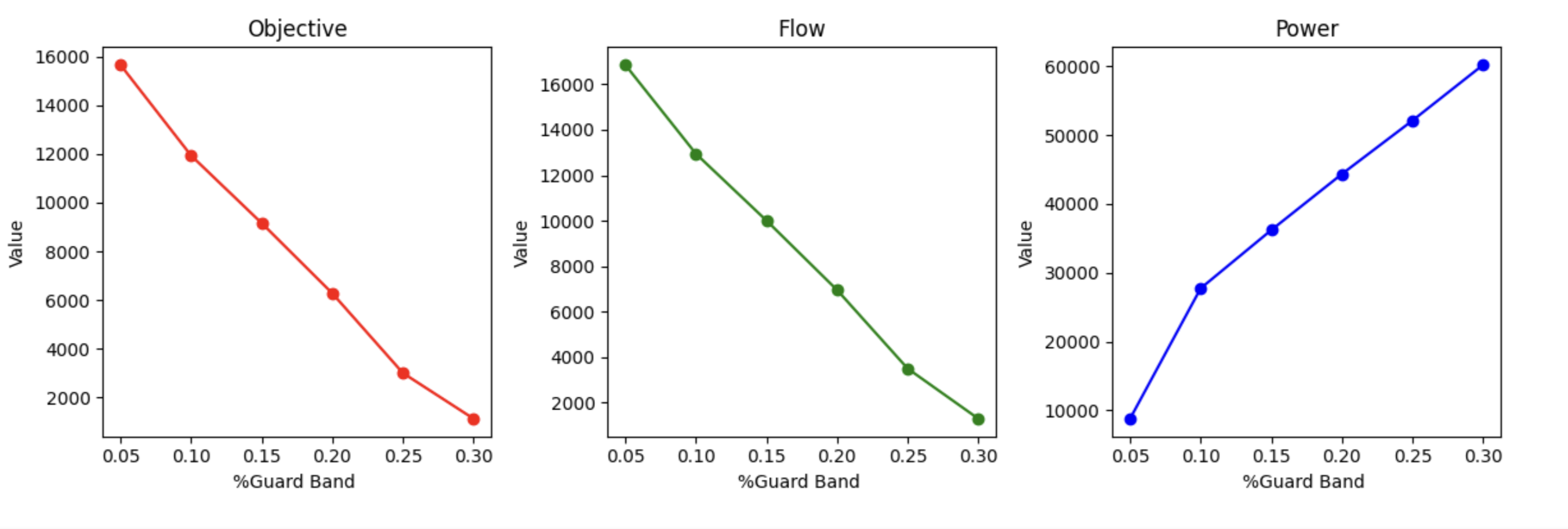}
    \caption{{\small The variation of the combined objective, the power and the flow, under different levels of inter-platoon safety.}}
    \label{fig:guard_band_exp}
     \vspace{-5pt}
\end{figure}

\begin{figure}
    \centering
    \includegraphics[width=1\linewidth]{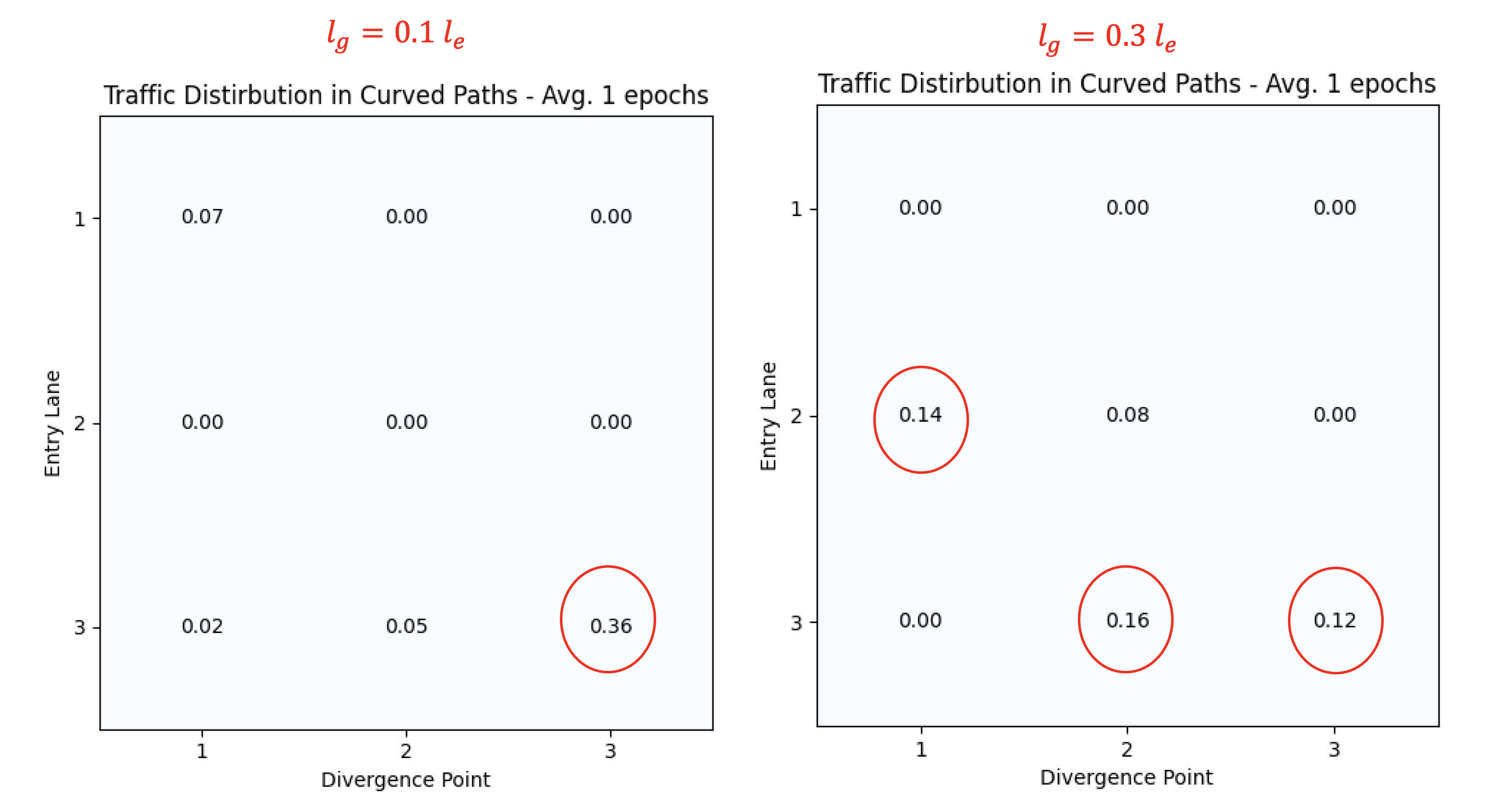}
    \caption{The marginal distribution of curved path assignment to complete left-turns. Left: low inter-platoon safety, more vehicles on shorter paths (lane 3), Right: a portion of vehicles shifting to longer paths due to lower capacity.Divergence point and Entry lanes are shown in Fig. \ref{fig:optimal_variables}}
    \label{fig:guard_band_dist}
\end{figure}

As previously mentioned, the guard band length ($l_g$) accounts for vehicle kinematic uncertainties affecting inter-platoon safety, which have a high-cost collision. On the other hand, increasing this guard band directly affects the segment capacity and therefore the intersection capacity. As seen in Fig. \ref{fig:guard_band_exp}, with increasing the guard band from $5\%$ to $30\%$ of a segment length around a virtual platoon, the objective and flow decrease, while the power consumption increases, confirming the expected trade-off between the objectives and safety. This behavior meets the expectations, as flow is expected to decrease as vehicles have to remain in the platoon from both ends and tend to acquire smaller velocity average values. Despite the decrease in average velocity, the increasing energy suggests vehicles take longer routes to complete a turn, which is a direct cause of lower capacity of entry lanes due to larger safety gaps. (Fig. \ref{fig:guard_band_dist})

\subsection{Straight-Left Demand Balance Analysis}

\begin{figure}
    \centering
    \includegraphics[width=0.7\linewidth]{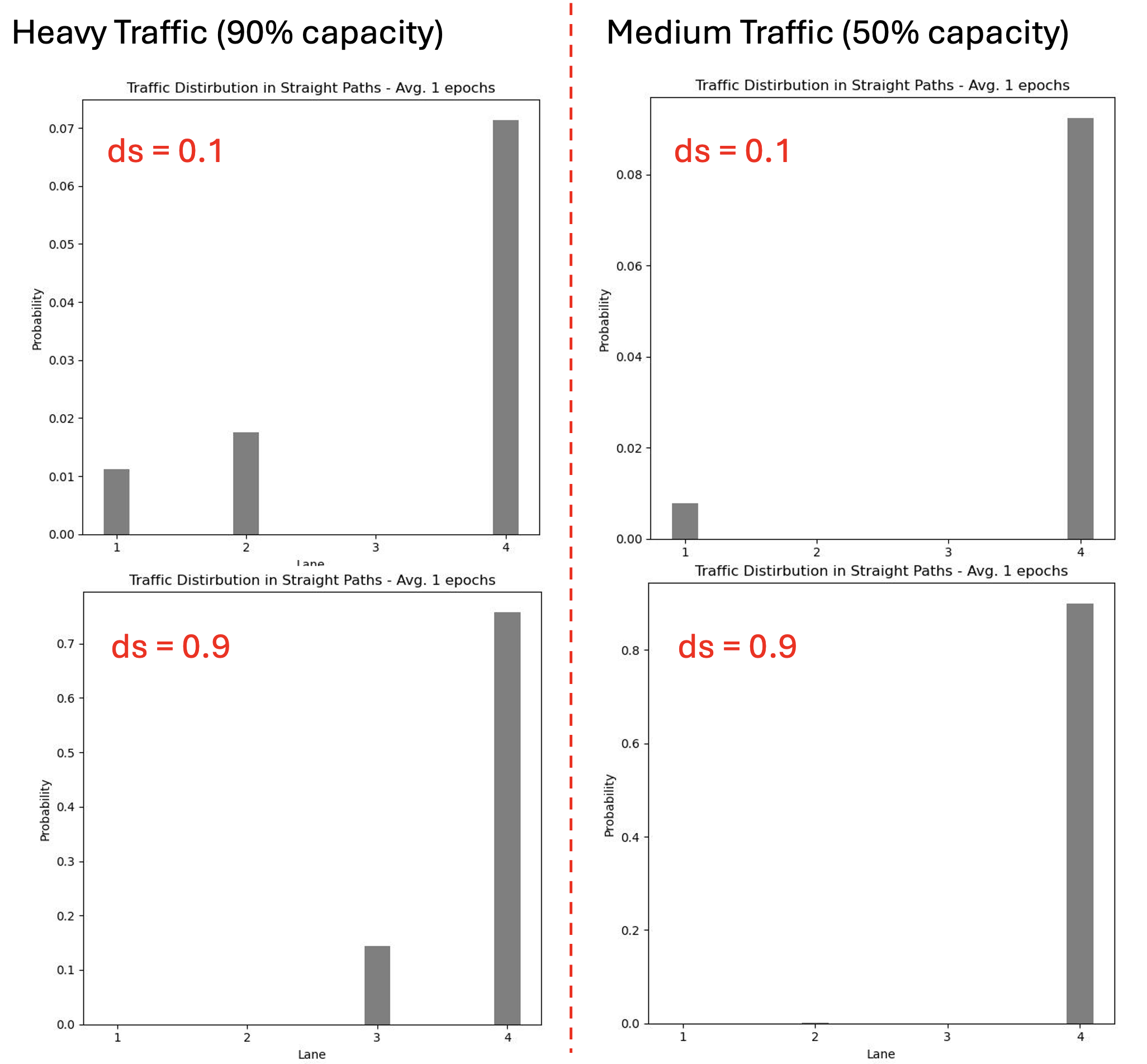}
    \caption{Straight traffic shift to shorter lanes in higher demands. In heavy traffic, the rightmost lane (lane 4) cannot guide all traffic as it reaches capacity.}
    \label{fig:demand_balance}
\end{figure}

There are two scenarios in which we examine the effect of the portion of straight/left traffic with a fixed total entry flow. A medium-traffic scenario (50\% entry capacity),and a heavy-traffic scenario. In general, there is flow-energy trade off in distributing straight traffic along lanes. Shorter paths have lower energy while longer paths have slightly higher average flow. It is observed that at $\alpha = 0.9845$ as the optimal weight, increasing $d_s$ (demand for straight compared to left) increases the overall flow and the objective while decreasing the energy, as straight paths are shorter. Moreover, straight path distribution is shifted towards in the leftmost lane to provide the shortest path (least energy), which is also not shared with left traffic. Yet, in heavy traffic scenarios, the left-most lane's flow reaches capacity and part of the straight traffic remains in other lanes. (Fig. \ref{fig:demand_balance})

\section{Conclusion}

In this work, we proposed a novel formulation of space and time averaged energy and flow at aerial intersections within a rhythmic control framework, where time arrivals are used as trajectory constraints. While minimizing average intersection power and maximizing intersection flow, the intersection demonstrated effectiveness under various magnitudes of entry flow (demand) and safety factors. The proposed traffic distribution and trajectory optimization approach can be served as a solution to guiding real-time UAM traffic at dense unsignalized aerial intersections,  paving the way for more integrated, energy-efficient, and high-performance urban aerial transportation systems.

\bibliographystyle{ieeetr}
\bibliography{main}


\addtolength{\textheight}{-12cm}   

\end{document}